\documentclass[prc,twocolumn,showpacs,preprintnumbers,amsmath,amssymb]{revtex4-1}
\usepackage[dvips]{graphicx}
\usepackage{bm}

\begin{document}
\title{Extended random-phase-approximation study of fragmentation of giant quadrupole resonance in $^{16}$O}
\author{Mitsuru Tohyama}
\affiliation{Faculty of Medicine, Kyorin University, Mitaka, Tokyo
  181-8611, Japan \email{tohyama@ks.kyorin-u.ac.jp}}
\begin{abstract}
The damping of isoscalar giant quadrupole resonance in $^{16}$O
is studied using extended
random-phase-approximation approaches derived from the time-dependent density-matrix theory. 
It is pointed out that the effects of ground-state correlations bring strong fragmentation of quadrupole strength even if the number of 
two particle--two hole configurations is strongly  limited. 
\end{abstract}
\maketitle
Giant resonances are considered to be highly collective states consisting of one particle (p) - one hole (h) excitations which can be treated in
the random phase approximation (RPA) based on the Hartree-Fock (HF) ground state. 
Most observed giant resonances show strong fragmentation of transition strength as is the case of the giant quadrupole resonance (GQR) in $^{16}$O \cite{lui}. 
This indicates that beyond RPA approaches which include higher configurations and also ground-state correlation effects are needed in realistic description of giant resonances. 
The second RPA (SRPA) \cite{srpa} which is based on the HF ground state includes the coupling of the 1p--1h states to 2p--2h states. 
In the particle-vibration coupling or quasiparticle-phonon models \cite{elena} p--h correlations among the 2p--2h configurations are expressed by phonons
and the effects of ground-state correlations are included in some versions of the particle-vibration coupling models \cite{kamer}.
Our extended RPA (ERPA) which is formulated by using the equation-of-motion approach \cite{s21} and a correlated ground state in the time-dependent density-matrix theory (TDDM) \cite{WC,GT,toh20}
includes both the coupling to higher configurations and the effects of ground-state correlations.
In ERPA the effects of ground-state correlations are included through the fractional occupation probability $n_\alpha$ of a single-particle state $\alpha$ 
and the correlated part $C_2$ of a two-body density matrix.
The small amplitude limit of TDDM (STDDM) which has been used for the study of giant resonances in oxygen and calcium isotopes \cite{toh07,toh18}
also includes $n_\alpha$ and $C_2$ but some correlation effects in two-body configurations space such as self-energy contributions are missing in STDDM.
In this paper ERPA is applied to GQR in $^{16}$O and the obtained results in ERPA are compared with those in STDDM.
It is shown that both ERPA and STDDM give the highly fragmented quadrupole strength in $^{16}$O \cite{lui} and that ERPA improves STDDM.
The correlations among the two-body configurations included in ERPA and STDDM have never been considered in the applications of other extended RPA theories 
that incorporate the effects of ground correlations through $n_\alpha$ and $C_2$ \cite{srpa,taka2,robin}. 
 
The correlated ground state used to formulate ERPA is given as a stationary solution of the TDDM equations. 
The TDDM equations consist of the coupled equations of motion for the one-body density matrix $n_{\alpha\alpha'}$ 
(the occupation matrix) and the correlated part of the two-body density matrix $C_{\alpha\beta\alpha'\beta'}$
($C_2$). In general the equations of motion for reduced density matrices form
a chain of coupled equations known as the Bogoliubov-Born-Green-Kirkwood-Yvon (BBGKY) hierarchy and $C_2$ couples to the correlated part $C_3$ of a three-body density matrix.
Approximations for $C_3$ are needed to close the equations of motion within $n_{\alpha\alpha'}$ and $C_2$. A few truncation schemes of the BBGKY hierarchy have so far been proposed
\cite{toh20}. In this work the simplest but reliable truncation scheme that neglects 
$C_3$ and takes only the 2p--2h and 2h--2p components of $C_2$ is used. It has been shown \cite{toh15} that this truncation scheme gives the ground state of
$^{16}$O which can be compared with the result in exact diagonalization approach (EDA).
The stationary solution of the TDDM equations can be obtained by using either an adiabatic method or a usual gradient method \cite{toh20}.

The ERPA equations for one-body amplitudes $x^\mu_{\alpha\alpha'}$ and two-body amplitudes $X^\mu_{\alpha\beta\alpha'\beta'}$ are derived from
the equation-of-motion approach \cite{s21} assuming the excitation operator 
\begin{eqnarray}
Q^\dag_\mu=\sum_{\alpha\alpha'}x^\mu_{\alpha\alpha'}a^\dag_\alpha a_{\alpha'}+\sum_{\alpha\beta\alpha'\beta'}X^\mu_{\alpha\beta\alpha'\beta'}a^\dag_\alpha a^\dag_\beta a_{\beta'} a_{\alpha'}
\end{eqnarray}
destructs the ground state $|0\rangle$ as $Q_\mu|0\rangle=0$ and excites an excited state $|\mu\rangle$ as $|\mu\rangle=Q^\dag_\mu|0\rangle$. 
Here, $a^\dag_\alpha ~(a_\alpha)$ is the creation (annihilation) operator of a nucleon at a single-particle state $\alpha$.
The equations in ERPA are written in the matrix form 
\begin{eqnarray}
\left(
\begin{array}{cc}
A&B\\
C&D
\end{array}
\right)\left(
\begin{array}{c}
{x}^\mu\\
{X}^\mu
\end{array}
\right)
=\omega_\mu
\left(
\begin{array}{cc}
S_{11}&T_{12}\\
T_{21}&S_{22}
\end{array}
\right)
\left(
\begin{array}{c}
{x}^\mu\\
{X}^\mu
\end{array}
\right),
\label{ERPA1}
\end{eqnarray}
where $\omega_\mu$ is the excitation energy of an excited state $|\mu\rangle$,
$A$, $B$, $C$ and $D$ are the ground-state expectation values of the double commutators between the Hamiltonian and either one-body or two-body excitation operators while 
$S_{11}$, $T_{12}~(=T_{21}^\dag)$ and $S_{22}$ are the ground-state expectation values of the commutators between either one-body or two-body excitation operators.
Each matrix element in Eq. (\ref{ERPA1}) is given explicitly in Ref. \cite{ts08}.
As mentioned above, the effects of ground-state correlations are included
in Eq. (\ref{ERPA1}) through $n_{\alpha}$ and $C_2$. 
The equation in SRPA is derived from Eq.(\ref{ERPA1}) by simply assuming the HF ground state where $n_\alpha$ is either 1 or 0 and
$C_2=0$.
The equations in STDDM can also be expressed in 
matrix form similar to Eq. (\ref{ERPA1}) but there is a difference in $D$: The matrix $D$ in STDDM does not contain the terms 
corresponding to the self-energy contributions to two-body configurations \cite{s21}.
Let us explain this point in more detail.
From the small amplitude limit of the TDDM equations that do not include $C_3$, the coupled equations in STDDM are obtained 
for the one-body transition amplitudes $\tilde{x}^\mu_{\alpha\alpha'}=\langle 0|a^\dag_{\alpha'}a_\alpha|\mu\rangle$ and the two-body transition amplitudes 
$\tilde{X}^\mu_{\alpha\beta\alpha'\beta'}=\langle 0|a^\dag_{\alpha'}a^\dag_{\beta'}a_\beta a_\alpha|\mu\rangle$. They are written in matrix form as
\begin{eqnarray}
\left(
\begin{array}{cc}
a&b\\
c&d
\end{array}
\right)\left(
\begin{array}{c}
\tilde{x}^\mu\\
\tilde{X}^\mu
\end{array}
\right)
=\omega_\mu
\left(
\begin{array}{c}
\tilde{x}^\mu\\
\tilde{X}^\mu
\end{array}
\right).
\label{STDDM}
\end{eqnarray}
The matrices $a$, $b$, $c$ and $d$ are also given in Ref. \cite{ts08}.
With the use of 
\begin{eqnarray} 
\left(
\begin{array}{c}
\tilde{x}^\mu\\
\tilde{X}^\mu
\end{array}
\right)=
\left(
\begin{array}{cc}
S_{11}&T_{12}\\
T_{21}&S_{22}
\end{array}
\right)
\left(
\begin{array}{c}
{x}^\mu\\
{X}^\mu
\end{array}
\right),
\end{eqnarray}
Eq. (\ref{STDDM}) can be transformed to another matrix form similar to Eq. (\ref{ERPA1}) as 
\begin{eqnarray}
\left(
\begin{array}{cc}
A&B\\
C&D'
\end{array}
\right)\left(
\begin{array}{c}
{x}^\mu\\
{X}^\mu
\end{array}
\right)
=\omega_\mu
\left(
\begin{array}{cc}
S_{11}&T_{12}\\
T_{21}&S_{22}
\end{array}
\right)
\left(
\begin{array}{c}
{x}^\mu\\
{X}^\mu
\end{array}
\right),
\label{STDDM1}
\end{eqnarray}
where $A=aS_{11}+bT_{21}$, $B=aT_{12}+bS_{22}$, $C=cS_{11}+dT_{21}$ and $D'=cT_{12}+dS_{22}$.
The matrices $A$, $B$ and $C$ are the same as those in Eq. (\ref{ERPA1}) but $D'\neq D$.
In order to express $D$ that is the ground-state expectation value of the double commutators between the Hamiltonian and two-body excitation operators,
an additional term $eT_{32}$ is needed, where $e$ depicts the coupling of the two-body transition amplitudes to the three-body transition amplitudes and $T_{32}$ is the expectation value of the commutator between
the three-body and two-body excitation operators \cite{ts08}. Thus ERPA includes the three-body effects that are not considered in STDDM. The terms in $eT_{32}$ express 
self-energy contributions and various vertex corrections \cite{s21}. One of the self-energy contributions to the 2p--2h configurations is written as 
\begin{eqnarray}
\Sigma({\rm p_1p_2h_1h_2:p_3p_4h_3h_4})&=&
-\delta_{\rm p_1p_3}\delta_{\rm {p_2p_4}}\delta_{\rm h_1h_3}
n_{\rm h_1}
\nonumber \\
&\times&\sum_{\rm pp'h}\langle {\rm pp'}|v|{\rm h}_2{\rm h}\rangle
C_{\rm h_4hpp'},
\label{self}
\end{eqnarray}
where $\langle {\rm pp'}|v|{\rm h}_2{\rm h}\rangle$ is the matrix element of the residual interaction $v$.

The single-particle states in the one-body and two-body amplitudes in Eqs. (\ref{ERPA1}) and (\ref{STDDM1}) can be both hole and particles states but
in the realistic applications of STDDM \cite{toh07,toh18}, $X^\mu_{\alpha\beta\alpha'\beta'}$ has been restricted to the 2p--2h and 2h--2p types to facilitate numerical calculations.
To investigate the validity of such a treatment of the two-body amplitudes, the fragmentation of the quadrupole strength in $^{16}$O is first studied by using
a small single-particle space consisting of the proton and neutron $1p_{1/2}$, $1p_{3/2}$ and $1d_{5/2}$ states for which the comparison with the EDA results can easily be made.
In this single-particle space no 1p--1h quadrupole transitions are allowed and the quadrupole strength can be carried by the 2p--2h configurations. 
Therefore, the comparison with the
EDA result tests the validity of ERPA and STDDM in the 2p--2h configuration space.
RPA and SRPA which are based on the HF ground state cannot give the quadrupole transitions in this single-particle space.
 
The single-particle energies and wavefunctions are calculated from the Skyrme III force \cite{skIII}.
A simplified interaction that contains only the $t_0$ and $t_3$ terms of the Skyrme III force is used as the residual 
interaction \cite{toh07}. It has been shown \cite{toh07} that the simple force induces ground-state correlations 
which are comparable to  the results of other theoretical calculations \cite{agassi,adachi,utsuno}.
The ground state is obtained by using the adiabatic method, which is explained in Ref. \cite{toh21} in some detail.
The 2p--2h and 2h--2p amplitudes in ERPA and STDDM are defined by using the same single-particle states as those used in the ground-state calculation.
The occupation probabilities calculated in TDDM are shown in Table \ref{tab1}. 
The results in EDA which are obtained by using the same single-particle states and interaction as those used in TDDM
are given in parentheses. The results in TDDM agree well with the EDA results.
The deviation of $n_\alpha$ from the HF values ($n_\alpha=1$ or 0) is close to 10 \%, 
indicating that the ground state of $^{16}$O is highly correlated.
The total energy in TDDM is 7.9 MeV lower than that in HF: The correlation energy is $-22.8 $MeV but it is largely compensated by the increase in the mean-field energy due
to the relaxation of the occupation probabilities from the HF values as explained in Ref. \cite{toh07}. As pointed out in Ref. \cite{toh21}, a few percent reduction of the Skyrme
parameters would be needed to reproduce the HF total energy in TDDM.
\begin{table}
\caption{Single-particle energies $\epsilon_\alpha$ and occupation probabilities 
$n_{\alpha}$ calculated in TDDM for $^{16}$O. The results in EDA are given in parentheses.}
\begin{center}
\begin{tabular}{c rr rr} \hline
 &\multicolumn{2}{c}{$\epsilon_\alpha$ [MeV]}&\multicolumn{2}{c}{$n_{\alpha}$}\\ \hline 
orbit & proton & neutron  & proton & neutron  \\ \hline
$1p_{3/2}$ & -18.2 & -21.8 & 0.907 (0.910) & 0.908 (0.910)  \\
$1p_{1/2}$ & -12.0 & -15.6 & 0.879 (0.883)  & 0.879 (0.883)  \\
$1d_{5/2}$ & -3.8 & -7.2 & 0.102 (0.099)  & 0.102 (0.099)   \\\hline
\end{tabular}
\label{tab1}
\end{center}
\end{table}

\begin{figure} 
\begin{center} 
\includegraphics[height=6cm]{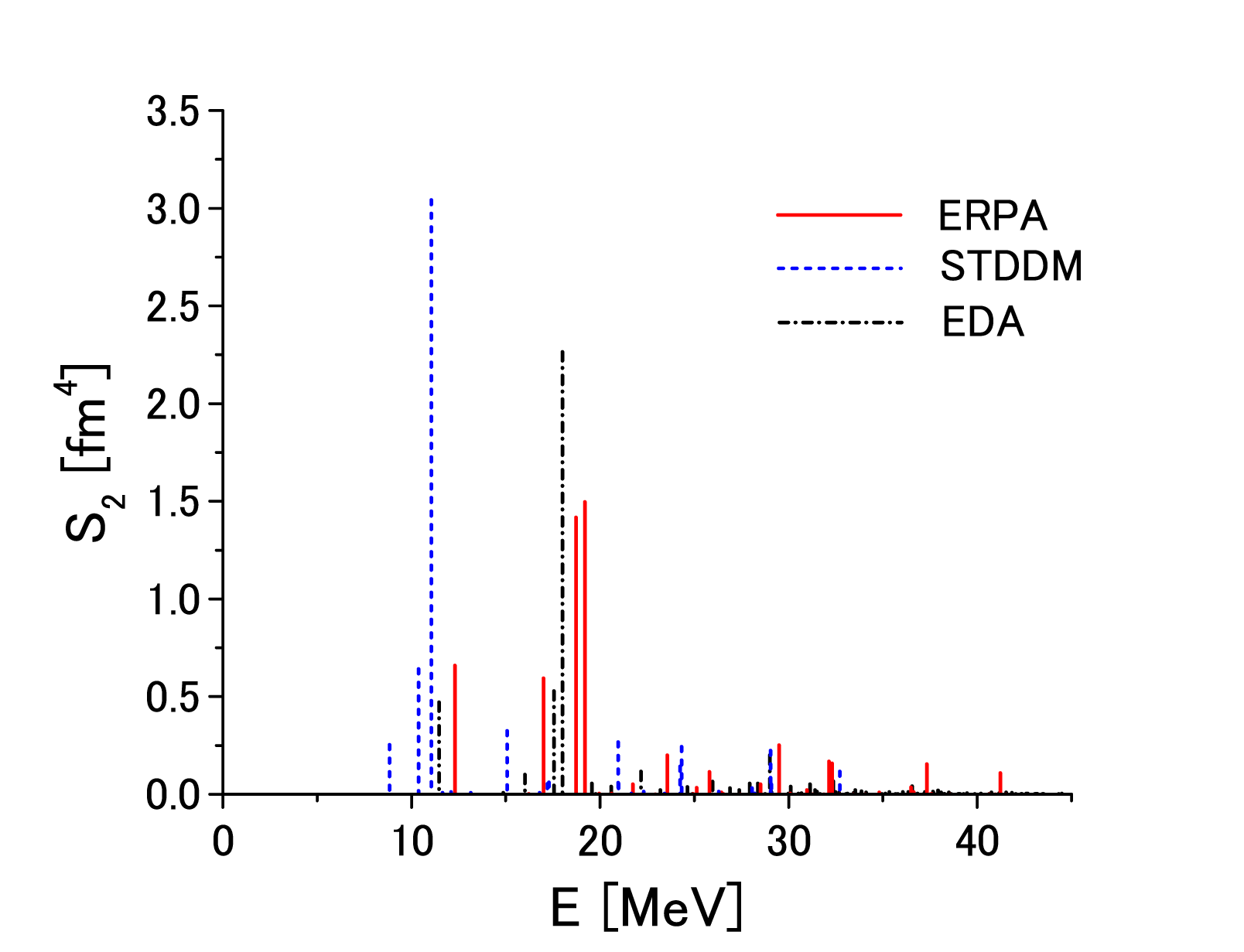}
\end{center}
\caption{Isoscalar quadrupole strength distributions calculated in ERPA (solid lines), STDDM (dotted lines)  and EDA (dot-dashed lines) for $^{16}$O.} 
\label{quad} 
\end{figure}

\begin{figure} 
\begin{center} 
\includegraphics[height=4cm]{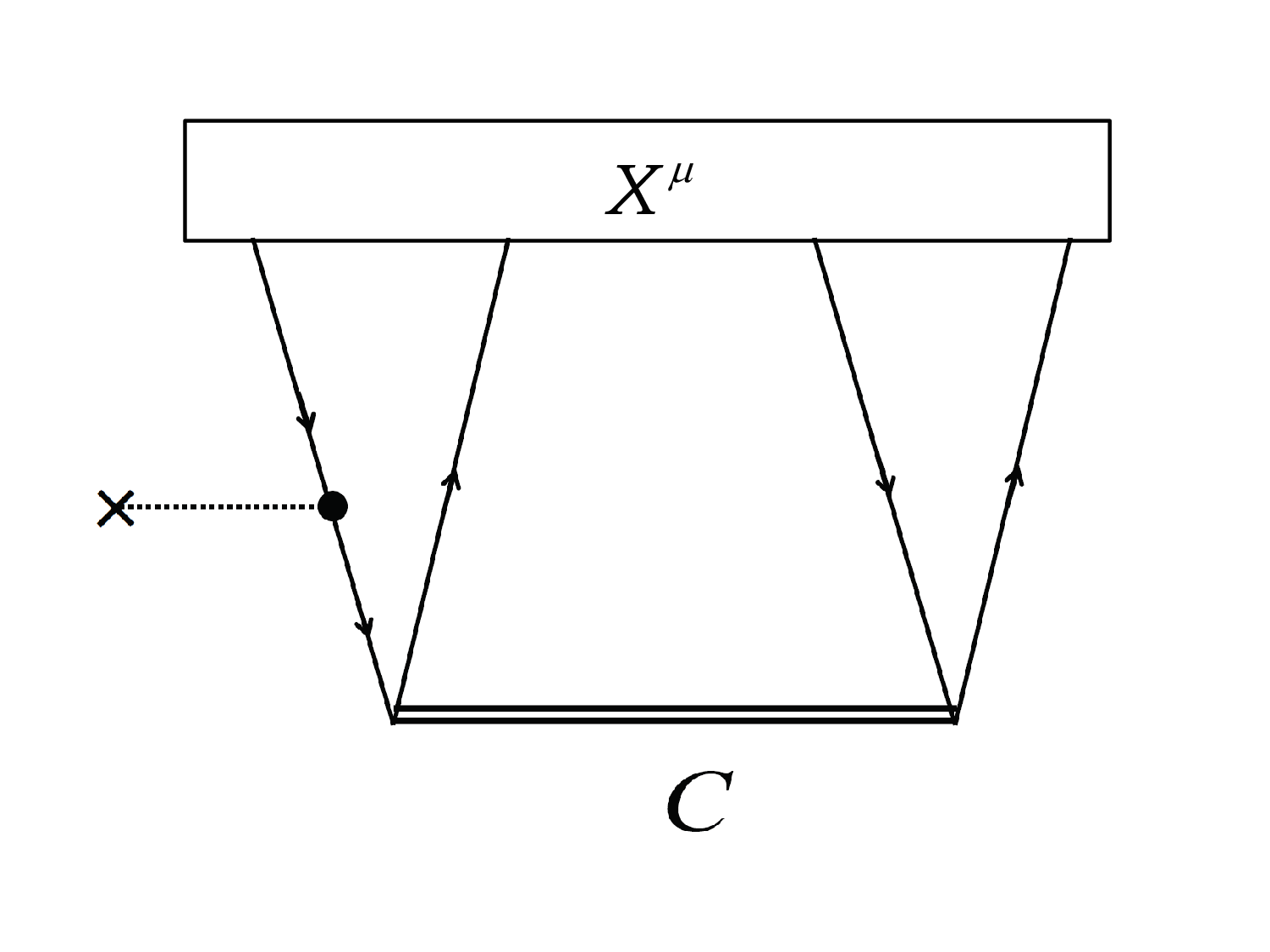}
\end{center}
\caption{Coupling of the h-h transition amplitude to the 2p-2h amplitude (square box) through $C_{\rm pp'hh'}$. 
The horizontal line indicates $C_{\rm pp'hh'}$
and the vertical lines with arrows either a hole state or a particle state. The dotted line with a cross at the left end depicts the
external field and the dot the 1h--1h transition amplitude.} 
\label{diag} 
\end{figure}

In Fig. \ref{quad} the isoscalar quadrupole strength distributions calculated in ERPA (solid lines), STDDM (dotted lines) and EDA (dot-dashed lines) are shown.
The quadrupole transition strengths in ERPA and STDDM are calculated from the one-body transition amplitude $\tilde{x}^\mu_{\alpha\alpha'}$ given by
\begin{eqnarray}
\tilde{x}^\mu_{\alpha\alpha'}=\langle 0|a^\dag_{\alpha'}a_{\alpha}|\mu\rangle=S_{11}x^\mu+T_{12}X^\mu.
\label{trans}
\end{eqnarray}
In the single-particle space used here the quadrupole transition strength is given by the second term and its process is depicted in Fig. \ref{diag}, 
where the horizontal line indicates $C_{\rm pp'hh'}$, the square box connected to the four vertical lines $X^\mu$, the dot $\tilde{x}_{hh'}^\mu$ 
and the dotted line with a cross at the left end the
external field. In SRPA the 2p--2h quadrupole configurations cannot make the one-body transition amplitudes because $T_{12}=0$ in Eq. (\ref{trans}).
The quadrupole strengths in ERPA and STDDM are largely fragmented, which agrees with the EDA result though there is some difference in the location and strength of each state.
The excitation energy of the main peak in ERPA is 8.1 MeV higher than that in STDDM and is close to the EDA result. 
This upward shift of the ERPA strength is explained by the self-energy contributions included in ERPA through $eT_{32}$.
Since the 1p--1h transitions are not allowed in the small single-particle space, the energy-weighted-sum-rule (EWSR) values exhausted by the quadrupole states shown in Fig. \ref{quad} are small: They are  6.7 \% , 4.5 \% 
and 5.7 \% in ERPA, STDDM and EDA, respectively.
Figure \ref{quad} clearly shows that ERPA reasonably well describes the correlations among the two-body configurations.

\begin{figure} 
\begin{center} 
\includegraphics[height=7cm]{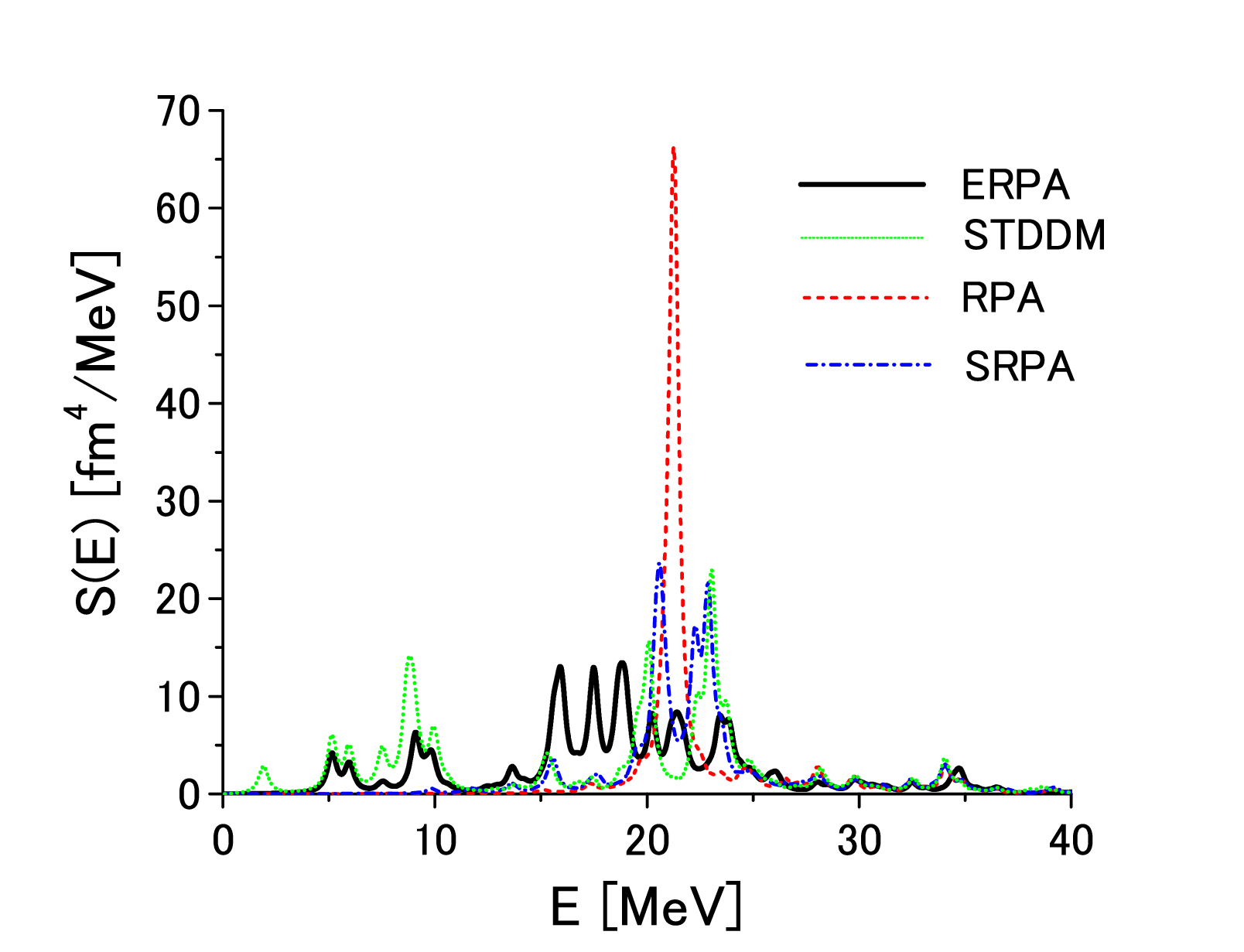}
\end{center}
\caption{Isoscalar quadrupole strength distributions calculated in ERPA (solid line), STDDM (dotted line), RPA (dashed line)and SRPA (dot-dashed line) for $^{16}$O.
The distributions are smoothed with an artificial width $\Gamma=0.5$ MeV.}
\label{oe2} 
\end{figure}

\begin{figure} 
\begin{center} 
\includegraphics[height=6cm]{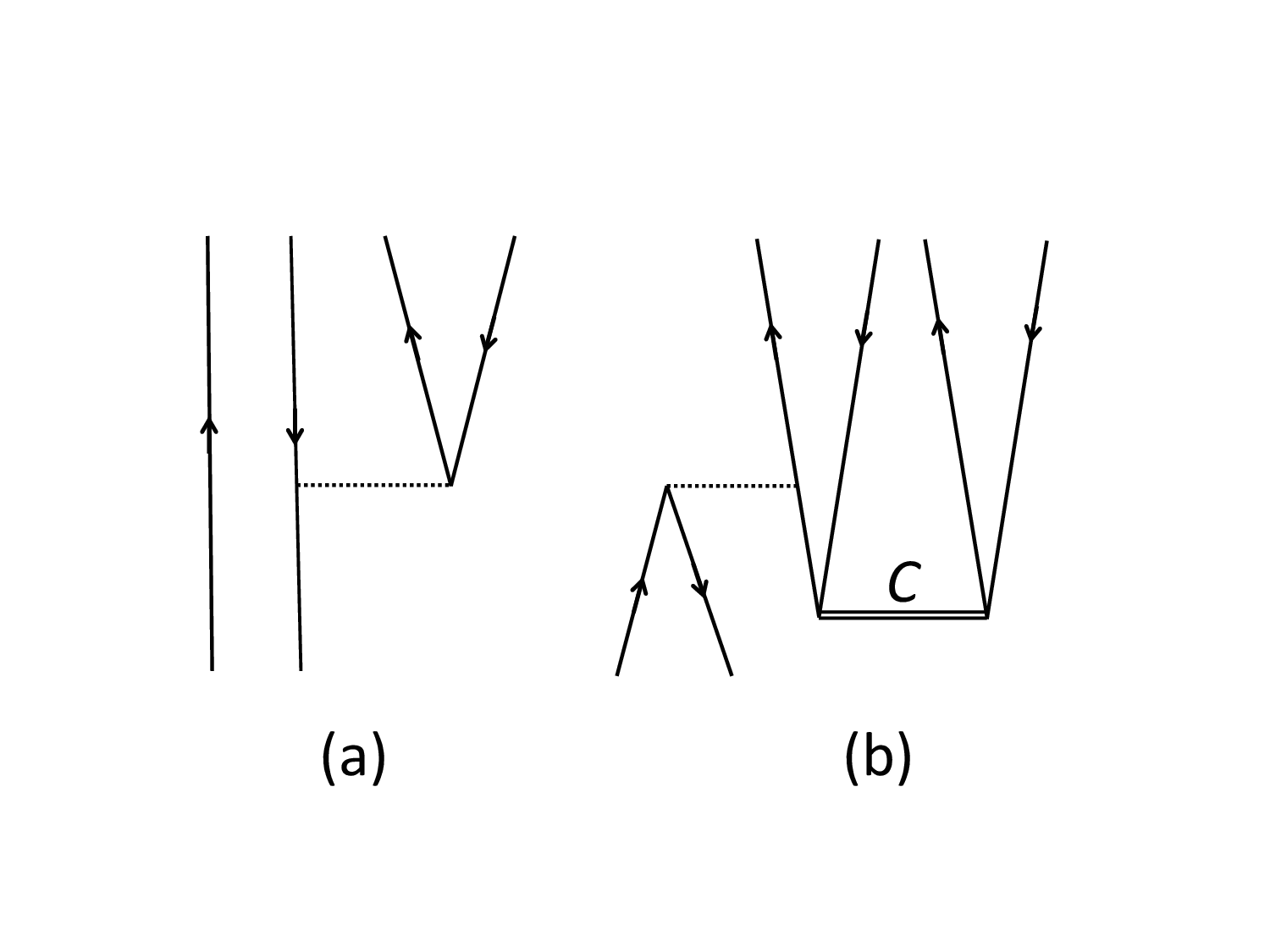}
\end{center}
\caption{Coupling of the 1p-1-h amplitude to the 2p--2h amplitude. 
The horizontal line indicates $C_{\rm pp'hh'}$, the two and four vertical 
lines with arrows $x^\mu_{\rm ph}$ and $X^\mu_{\rm pp'hh'}$, respectively, and the dotted line the residual interaction.
ERPA and STDDM include both processed (a) and (b) while SRPA includes only processes (a).} 
\label{diag2} 
\end{figure} 

\begin{figure} 
\begin{center} 
\includegraphics[height=6cm]{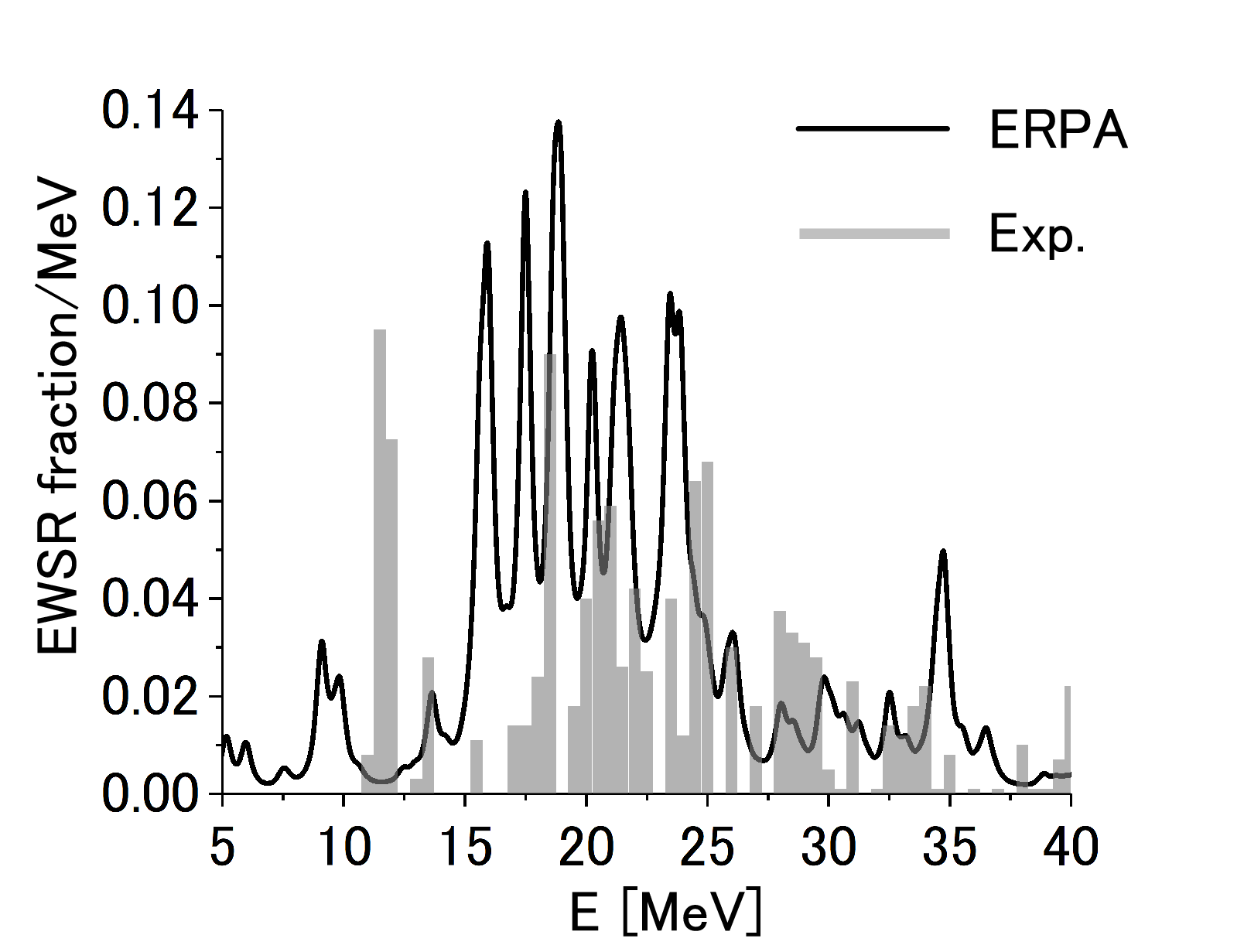}
\end{center}
\caption{Isoscalar quadrupole strength distribution calculated in ERPA (solid line) for $^{16}$O
is shown as a ratio to the EWSR value and compared with
the observed quadrupole strengths above 10 MeV  \cite{lui} (vertical bars).}
\label{oe21} 
\end{figure}

The results of realistic ERPA and STDDM calculations that include a large number of single-particle states for $x^\mu_{\alpha\alpha'}$ and thus can be compared with experiment
are presented below.
The one-body amplitudes ${x}^\mu_{\alpha\alpha'}$ are defined with a large number of single-particle states including those in the 
continuum: The continuum states are discretized by confining the wavefunctions in a sphere with radius 15 fm and all 
the single-particle states with $\epsilon_\alpha\le 50$ MeV and 
$j_\alpha\le 9/2 \hbar$ are included. 
Since the simple residual interaction which is also used here is not consistent with the full Skyrme III interaction,
it is necessary to reduce the strength of the residual interaction in the one-body channels when the large single-particle space is used for ${x}^\mu_{\alpha\alpha'}$. 
The reduction factor $f$ is determined so that the spurious mode corresponding
to the center-of-mass motion comes at zero excitation energy in RPA. It is found that $f=0.62$. This factor is used in the $A$, $B$ and $C$ parts of Eqs. (\ref{ERPA1}) and (\ref{STDDM1}).
The 2p--2h and 2h--2p amplitudes are defined by using the same single-particle states as those used in the ground-state calculation. 
The results of the ERPA and STDDM calculations for the isoscalar quadrupole excitation in $^{16}$O are shown in Fig. \ref{oe2}
with the solid and dotted lines, respectively. The dashed and dot-dashed lines depict the results in RPA and SRPA, respectively. 
The distributions are smoothed with an artificial width $\Gamma=0.5$ MeV.
The peak in RPA corresponds to GQR.
The EWSR values exhausted by RPA and SRPA are 104 \% while those in ERPA and STDDM are 106 \% and 105 \%, respectively.
To fulfill EWSR completely, the self-consistent use of the residual interaction and the better treatment of the continuum states would be needed. 
Due to the couping to the 2p-2h configurations GQR is fragmented in SRPA, ERPA and STDDM. However, the SRPA result has no visible strength below 10 MeV.
Figure \ref{oe2} shows that the inclusion of the ground-state correlations significantly increases the fragmentation of the quadrupole strength below GQR.
In ERPA and STDDM the coupling of the 1p--1h amplitudes to the 2p--2h amplitudes is enhanced due to the ground-state correlations through the process
depicted in Fig. \ref{diag2} (b). SRPA only includes the processes shown in Fig. \ref{diag2} (a).
The two peaks seen around 5.5 MeV in the STDDM and ERPA results are due to the 1p--1p transitions from the partially occupied $1d_{5/2}$ states.
The bumps between 7 MeV and 11 MeV in STDDM and ERPA are due mainly to the transitions from the 2p--2h configurations as depicted in Fig. \ref{diag}. These states correspond to the low-lying states seen
in Fig. \ref{quad}.
The small peak at 2 MeV in the STDDM result mainly consists of the 2p--2h configurations. It disappears in
ERPA because the self-energy contributions (Eq. (\ref{self})) push it into higher energy regions as is the case of the STDDM and ERPA calculations shown in Fig. \ref{quad}. 
The quadrupole strength in ERPA is more strongly fragmented between 12 MeV and 25 MeV than that in STDDM. This is another effect of
the $eT_{32}$ term that enhances the correlations among the 2p--2h configurations, though it is difficult to show which term in $eT_{32}$ is most important.
The large fragmentation in ERPA is comparable to the quadrupole strength distribution observed above 10 MeV \cite{lui} as shown in Fig. \ref{oe21}
though ERPA cannot fully reproduce the position and height of each peak.
Figure \ref{oe21} depicts the ratio to the EWSR value. The quadrupole strength observed between $E=11$ MeV and 40 MeV accounts for $53\pm10$ \% of EWSR \cite{lui}. 

In summary, the damping of isoscalar giant quadrupole resonance in $^{16}$O
was studied by using beyond RPA approaches, the extended RPA (ERPA) and the small amplitude of the time-dependent density-matrix theory (STDDM)
both derived from TDDM.
It was found that the effects of ground-state correlations bring strong fragmentation of quadrupole strength even in a small single-particle space used for 
two particle--two hole configurations. It was pointed out that self-energy contributions included in ERPA improve the results in STDDM.


\begin{thebibliography}{00}
\bibitem{lui}
Y.-W. Lui, H. L. Clark, and D. H. Youngblood, Phys. Rev. C 64, 064308 (2001).
\bibitem{srpa}
S. Dro$\dot{\rm z}$d$\dot{\rm z}$, S. Nishizaki, J. Speth, and J. Wambach, Phys. Rep. 197, 1 (1990).
\bibitem{elena}
E. Litvinova and P. Schuck, Phys. Rev. C {100}, 064320 (2019) and references therein.
\bibitem{kamer}
S. Kamerdzhiev, J. Speth, and G. Tertychny, Phys. Rep. 393, 1 (2004) and references therein.
\bibitem{s21}
P. Schuck, D. S. Delion, J. Dukelsky, M. Jemai, E. Litvinova, G. R$\ddot{\rm o}$pke, and M. Tohyama,
Phys. Rep. 929, 1 (2021).
\bibitem{WC}
S. J. Wang and W. Cassing, { Ann. Phys.} {159}, 328 (1985).
\bibitem{GT}
M. Gong and M. Tohyama, { Z. Phys. A} {335}, 153 (1990).
\bibitem{toh20}
M. Tohyama,  Front. Phys. {\bf 8}, 67 (2020).
\bibitem{toh07}
M. Tohyama, { Phys. Rev. C} 75, 044310 (2007).
\bibitem{toh18}
M. Tohyama, Prog. Theor. Exp. Phys. {2018}, 043D02 (2018).
\bibitem{taka2}
K. Takayanagi, K. Shimizu, A. Arima,
{Nucl. Phys. A} {\bf 481}, 313 (1988).\\
\bibitem{robin}
C. Robin C, E. Litvinova,
{\ Phys. Rev. Lett}. {\bf 123}, 202501(2019).\\
\bibitem{toh15}
M. Tohyama, Phys. Rev. C {\bf 91}, 017301 (2015).
\bibitem{ts08}
M. Tohyama and P. Schuck, Eur. Phys. J. A {36}, 349 (2008).
\bibitem{skIII}
M. Beiner, H. Flocard, Nguyen Van Giai, and P. Quentin.
{Nucl. Phys. A} {238}, 29 (1975).
\bibitem{agassi}
D. Agassi, V. Gillet, and A. Lumbroso, { Nucl. Phys.} A {130}, 129 (1969).
\bibitem{adachi}
S. Adachi, E. Lipparini, and Nguyen van Giai,
{ Nucl. Phys. A} {438}, 1 (1985).
\bibitem{utsuno}
Y. Utsuno and S. Chiba, Phys. Rev. C {\bf 83}, 021301 (2011).
\bibitem{toh21}
M. Tohyama, Prog. Theor. Exp. Phys. {2021}, 083D01 (2021).
\end{thebibliography}
\end{document}